\documentclass[a4paper,11pt]{article}
\pdfoutput=1 

\usepackage{jinstpub} 
\usepackage[]{subcaption}
\usepackage{amsmath}
\usepackage{array}
\usepackage{lineno}

\newcommand{\geff}{G$^{Eff}$}
\newcommand{\gtot}{G$^{Tot}$}
\newcommand{\ibf}{$IBF$}
\newcommand{\gphtot}{G$^{Tot}_{Ph}$}

\newcommand{\ngas}{n$^{gas}$}
\newcommand{\nliq}{n$^{liq}$}

\title{Charge and Light Production in the Charge Readout System of a Dual Phase LAr TPC}

\author[a]{T. Lux,\note{Corresponding author.}}

\affiliation[a]{Institut de F\'{i}sica d\'{}Altes Energies (IFAE) - Barcelona Institute of Science and Technology (BIST), Bellaterra (Barcelona), Spain}

\emailAdd{Thorsten.Lux@ifae.es}

\abstract{
For the future neutrino oscillation experiment DUNE, liquid argon time projection chambers with a fiducial mass of 10 kton each
are foreseen. The dual phase concept is one of the two implementations considered, wherein electrons produced by ionization in the liquid are extracted to a gaseous region above the liquid where they are amplified. For the amplification, large electron multipliers will be used. The technology was tested in various prototypes, most recently with a 3 x 1 x 1 m$^3$ large setup. 
An even larger prototype of 6 x 6 x 6 m$^3$ is currently being constructed and will start operation in 2019. An intensive 
R\&D program was carried out with the focus on achieving an effective gain of at least 20. In the simulation study here presented for the first time not only the electron signal is considered but also the ion backflow and the expected production of secondary scintillation light is studied, because the latter might limit the capability of the detector to trigger on low energetic
no-beam physics. It is found that the ion backflow and the light yield can be expected to be very large. The results
for the effective gain show a discrepancy with experimental data, both in size and shape of the gain curve. Based on literature studies, it is argued that photon feedback contributes to the gain in detectors filled with pure noble gases, especially in the case of pure argon. }


\arxivnumber{1812.08700} 

\begin{document}
\maketitle
\flushbottom

\section{Introduction}
For the next generation long baseline neutrino oscillation experiment DUNE (Deep Underground Neutrino Experiment) \cite{Acciarri:2016crz} Liquid Argon (LAr) based Time Projection Chambers (TPC) are the baseline choice for the far detectors. In total 4 LAr TPCs, each with a fiducial mass of 10 ktons, will be constructed until 2030. LAr has the advantage of providing the required large detector masses at
reasonable costs and of allowing the construction of a fully active tracking calorimeter. Two TPC concepts are currently under discussion for these detectors: a single phase (SP) and a dual phase (DP) LAr TPC concept. The SP LAr TPC concept was already used in previous experiments \cite{ARNEODO199795} and its implementation for the DUNE case is described elsewhere \cite{Cavanna:2271786}. The basic principle of a LAr DP TPC, and to a large extent also of any SP TPC,  is the following: A charged particle traversing the
LAr ionizes the argon atoms creating a large amount of electron-ion pairs on its way through the gas, a minimum ionizing particle creates in 1 cm about 50.000 electron-ion pairs. By applying an external electric field, the drift field, the electrons and ions
are separated and drift to the anode and cathode respectively. The field strength of the drift field is of the order of 0.5 to 1 kV/cm. While the ions are simply neutralized at the cathode, the electrons are detected at the anode to reconstruct the event. 
In a DP detector the electrons are extracted from the liquid to a gaseous volume by means of a strong electric field of 2 to 3 kV/cm and undergo afterwards a charge amplification before the signal is read out at the segmented anode. 
The DP concept has several potential advantages over the one of the SP: LAr is critically sensitive to oxygen impurities which causes the loss of electrons while drifting to the anode. Due to the amplification in the gas phase, these losses can be partly compensated
allowing lower energy detection thresholds, simpler electronics and longer drift distances, desirable to maximize the fiducial volume. The signals recorded at the segmented anode allow to reconstruct the trajectory of the charged particle in 3D. 
However, the signal measured at the anode does not provide information about the absolute $z$ coordinate, the position between the anode and the point where the event interaction took place. This information can be obtained from the primary scintillation
light (S1) which is emitted isotropically during the ionization process. Measuring the time difference between this light
pulse and the arrival of the electrons at the anode provides the $z$ coordinate, assuming the knowledge of the electron drift velocity. The capability to measure S1 is especially relevant to trigger on low energetic no-beam
physics events such as neutrinos emitted in a nearby Supernova explosion. The detection of such an event would not only
allow to understand the processes going on in the core of an exploding star but also could answer fundamental
questions in neutrino physics such as which mass hierarchy is realized in nature \cite{Migenda:2018ljh}.\\
The different technological aspects of the DP detector were developed over the years \cite{4774662,  1748-0221-9-03-P03017, 1742-6596-308-1-012027, BADERTSCHER2010188, 1757-899X-101-1-012049, 1748-0221-12-03-P03021, CALVO2018186} 
and studies are ongoing about the scalability of this technology towards the size needed for the DUNE far detector. A 4 tonne demonstrator was successfully constructed and operated in 2017 \cite{Aimard:2018yxp} 
and a 300 tonne prototype is currently being constructed starting operation in 2018 \cite{DeBonis:1692375}.\\
However, while significant efforts were undertaken to optimize the charge amplification region as a function of the charge signal detected at the anode, the aspect of secondary scintillation production in this region still needs further studies. Secondary scintillation (S2) is produced when an electron is accelerated in an electric field sufficiently high so that the electrons can obtain enough energy between two collisions to excite the gas atoms.
The process was studied intensively for the case of homogeneous electric fields \cite{MONTEIRO2008167} in the framework
of dark matter experiments. Studies about S2 production in micropattern gas detectors (MPGDs)  were also performed, experimentally \cite{MONTEIRO201218} as also by means of simulations \cite{1748-0221-7-09-P09006}. However, the performance depends strongly
on the exact geometry and field configuration, so that these studies can only serve as guidance. \\
The S2 light might affect the detection threshold important for the physics reach  in respect
to no-beam physics, a simulation study was carried out using well established Monte Carlo software. This study
is the first complete simulation, including charge, electrons and ions, as well as the S2 photons, of the charge readout region of a LAr DP TPC. 

\section{Simulation Framework}
For the simulations, a combination of several open source programs was used. The workflow follows the example given in \cite{Renner}:\\ 
Gmsh is a free 3D generator for finite element meshes \cite{2009IJNME..79.1309G}. The program contains a CAD engine and allows either to import geometries created by external CAD programs or to create geometries describing them by a boundary representation (BRep). The built-in scripting language allows to parametrize geometries. The mesh generated by Gmsh is imported by Elmer,  an open source to simulate multi-physics problems \cite{Elmer}. Among others, it includes physical models of fluid dynamics, structural mechanics, electromagnetic phenomena, heat transfer and acoustics. These are described by partial differential equations which Elmer solves by the Finite Element Method (FEM). In this study, it is used to produce the electric field map. Garfield++ is a C++ framework \cite{Veenhof:1993hz} developed for the simulation of electron transport in gas detectors of arbitrary geometry using Monte Carlo (MC) methods. Beside an interface to import field maps provided by external Finite Element Method (FEM) programs including Elmer, it is also interfaced with Magboltz \cite{Biagi:1999nwa}. Magboltz is an MC program developed to calculate the transport parameters of electrons drifting in gases and gas mixtures under the influence of electric and magnetic fields. For this purpose, the program contains a database with the electron cross sections and excitation and ionization energy levels for 60 gases. Garfield++ and Magboltz are well-established simulation tools which were used in a series of studies (e.g. \cite{506661,OLIVEIRA20131, Assran:2011ug}. 

\section{Analysis Strategy}
\begin{figure}[ht] 
\begin{center} 
\includegraphics[width=0.7\textwidth]{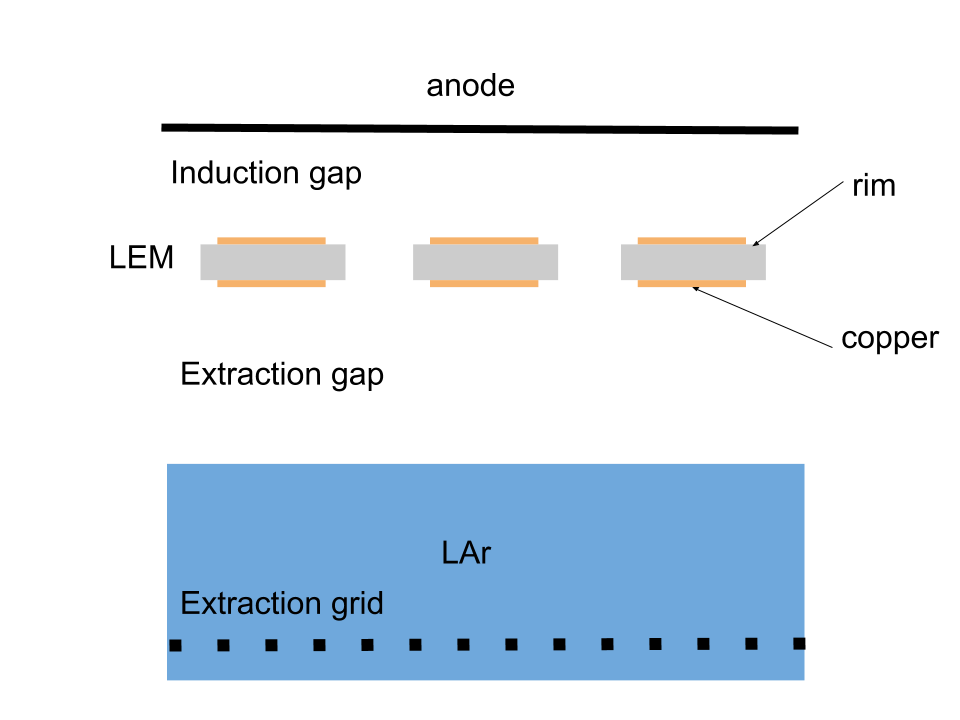} 
\caption{Sketch of the simulated geometry. The liquid is not included in the simulation.} \label{fig:geom}
\end{center} 
\end{figure}

The simulated geometry follows closely the one used in the prototypes, which is the baseline choice for DUNE.  The extraction and induction fields are fixed to their nominal values, 3 kV/cm, determined by the extraction efficiency, for the extraction field and 5 kV/cm for the induction field. The data were simulated for 90 K and 760 Torr. The amplification in the gaseous phase is achieved by using a Large Electron Multiplier (LEM), also referred to as Thick GEM (THGEM). A LEM consists of a printed circuit board, thickness usually between 0.4 and 1 mm, coated on both sides with copper, about 35 $\mu$m thick, in which high-density holes are drilled, typically ten thousands per device. The hole in the copper layers are normally slightly larger than the hole in the dielectric. This additional hole radius in the copper layer is called {\it rim}. Rim values up to 100 $\mu$m are common. The parameters describing the geometry and the electric field configuration used in the simulation are summarized in Tab.~\ref{tab:geomfield}. The simulated geometry is sketched in Fig.~\ref{fig:geom}. In the simulation only the gaseous phase is included. For each LEM voltage setting between 500 and 1000 events were simulated, for higher LEM voltages more events were simulated to reduce the statistical errors. Charging up effects were not included in the study and the presented results are compared to experimental data obtained before charging up effects become relevant. The simulation of the charging up effects in a THGEM are discussed in detail in \cite{Correia_2018}.
\begin{table}[htbp]
\begin{center}
\begin{tabular}{|c|c|}
\hline
\textbf{Geometry}   & \textbf{} \\
\hline
\textbf{Extraction gap}             & 5 mm  \\
\textbf{Induction gap}             & 2 mm   \\
\textbf{LEM dielectric thickness}             & 1 mm        \\ 
\textbf{LEM copper thickness}             & 35 $\mu$m        \\ 
\textbf{LEM dielectric hole radius}             & 250 $\mu$m        \\ 
\textbf{LEM copper hole rim}             & 40 $\mu$m         \\ 
\textbf{LEM hole pitch}             & 800 $\mu$m        \\ 
\textbf{LEM hole arrangement}             & hexagonal        \\ 
\hline
\textbf{Field/Voltage}   & \textbf{} \\
\hline
\textbf{Extraction field}             & 3 kV/cm   \\
\textbf{Induction field}             & 5 kV/cm   \\
\textbf{LEM Voltage}             & 2500 to 3500 V   \\
\hline
\end{tabular}
\caption{Summary of the geometry and the electric fields and voltages. \label{tab:geomfield}}
\end{center}
\end{table}

\begin{figure}[ht]
\begin{center}
\includegraphics[width=0.8\textwidth]{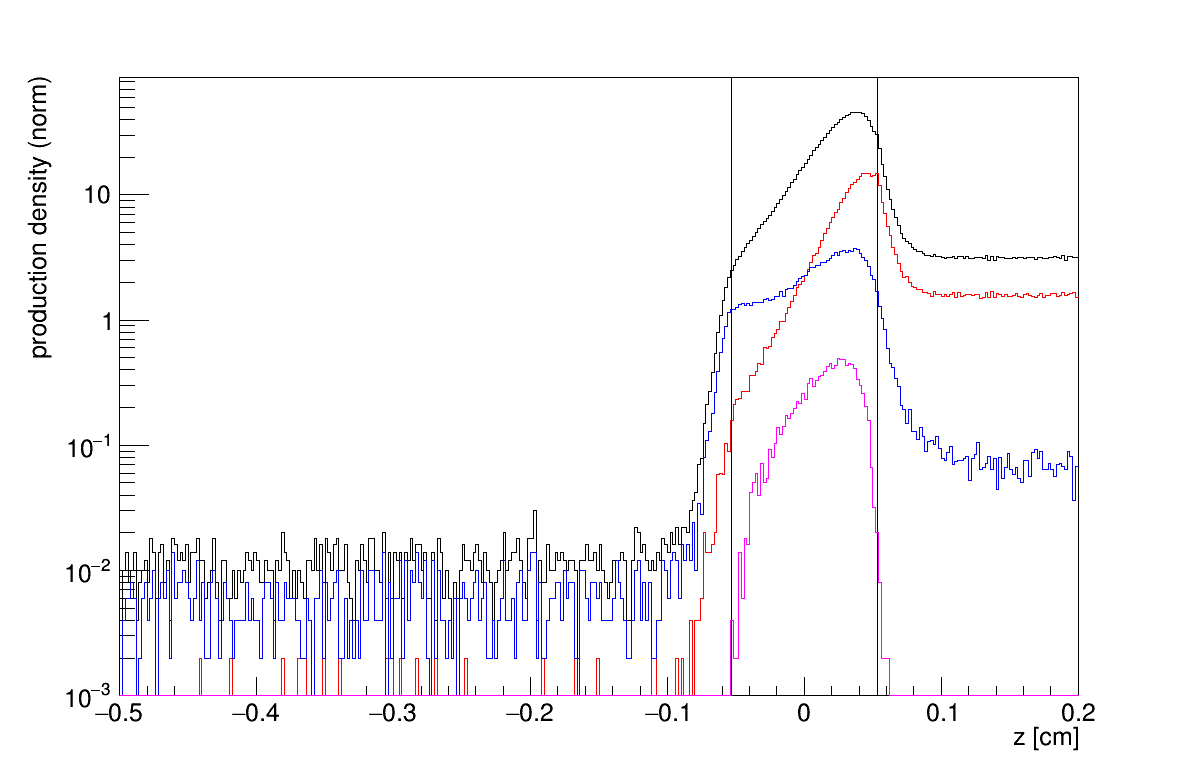}
\caption{$z$ distribution of the production point of S2 photons (black) and electron/ion paris (magenta) normalized to one event. The red curve shows the production point of the photons reaching the anode an in blue for the photons reaching the liquid surface. The vertical lines indicate the end of the position of the surfaces of the copper layers of the LEM. The anode is situated at 0.2 mm from the center of the LEM (at 0 cm). }
\label{fig:ZDistri}
\end{center}
\end{figure}

The primary electron was placed 0.49 mm ($z$ axis) below the bottom side of the LEM. Its starting position within the $xy$ plane, parallel to the liquid surface, was randomized. The electron was then transported using the microscopic avalanche MC method implemented in Garfield++ according to the local electric field provided by Elmer and the corresponding transport parameters from Magboltz. The position of each ionization as well as the location where the electrons and ions end their drift was stored in a ROOT file \cite{Brun:1997pa} for further analysis. Accordingly, the position of each excitation was retrieved following the approach developed in \cite{OLIVEIRA2011217}. As in that study, it is assumed that each excitation leads to the emission of a photon with a wavelength of 128 $\pm$ 10 nm \cite{Tanaka:55}. The direction of the photon was randomized in 4$\pi$. In a second step, the stored data was used for an analysis based on Python and ROOT. For the charge carriers, electrons and ions, the total gain, \gtot, the effective gain, \geff\ and the relative ion backflow, \ibf, were calculated. These variables are defined in the following way:
\begin{align*} 
G^{Tot} & =  \frac{\#\ \text{electrons produced}}{\#\ \text{primary electrons}} \\ 
G^{Eff} & =  \frac{\#\ \text{electrons reaching the anode}}{\#\ \text{primary electrons}} \\
IBF & = \frac{\#\ \text{ions reaching the liquid}}{\#\ \text{electrons reaching the anode}} 
\end{align*}
For the photons, Garfield++ only provides the production point and the transport of the photon had to be simulated separately. The endpoint of each photon was determined analytically and it was counted how many photons ended on the anode, the top side of the LEM, the bottom side of the LEM, escaped to the liquid or were absorbed by the dielectric of the LEM hole. \\
Fig.~\ref{fig:ZDistri} shows the production point along the $z$ axis for electron/ion pairs and photons for a LEM voltage of 3300 V, the nominal voltage setting at which an effective gain of 20 is expected based on experimental data \cite{Cantini:2014xza}. As one can see significantly more photons than electron/ion pairs are produced and most of the interactions happen close to the hole end towards the anode. While the ionization is restricted to the LEM holes, photons are also produced by excitation in significant quantities in the extraction and especially the induction gap.

\begin{figure}
\begin{subfigure}[b]{.5\linewidth}
\includegraphics[width=0.96\textwidth]{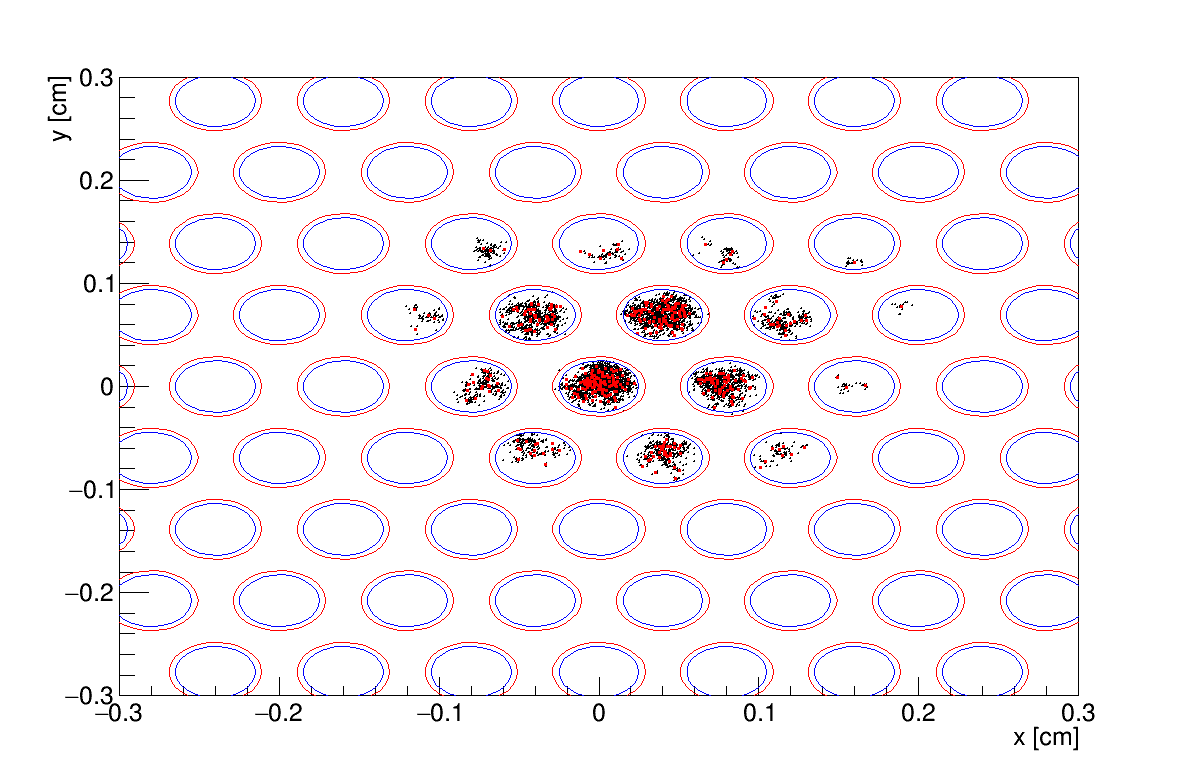}
\centering
\caption{Charge}\label{fig:XYCharge}
\end{subfigure}%
\begin{subfigure}[b]{.5\linewidth}
\centering
\includegraphics[width=0.96\textwidth]{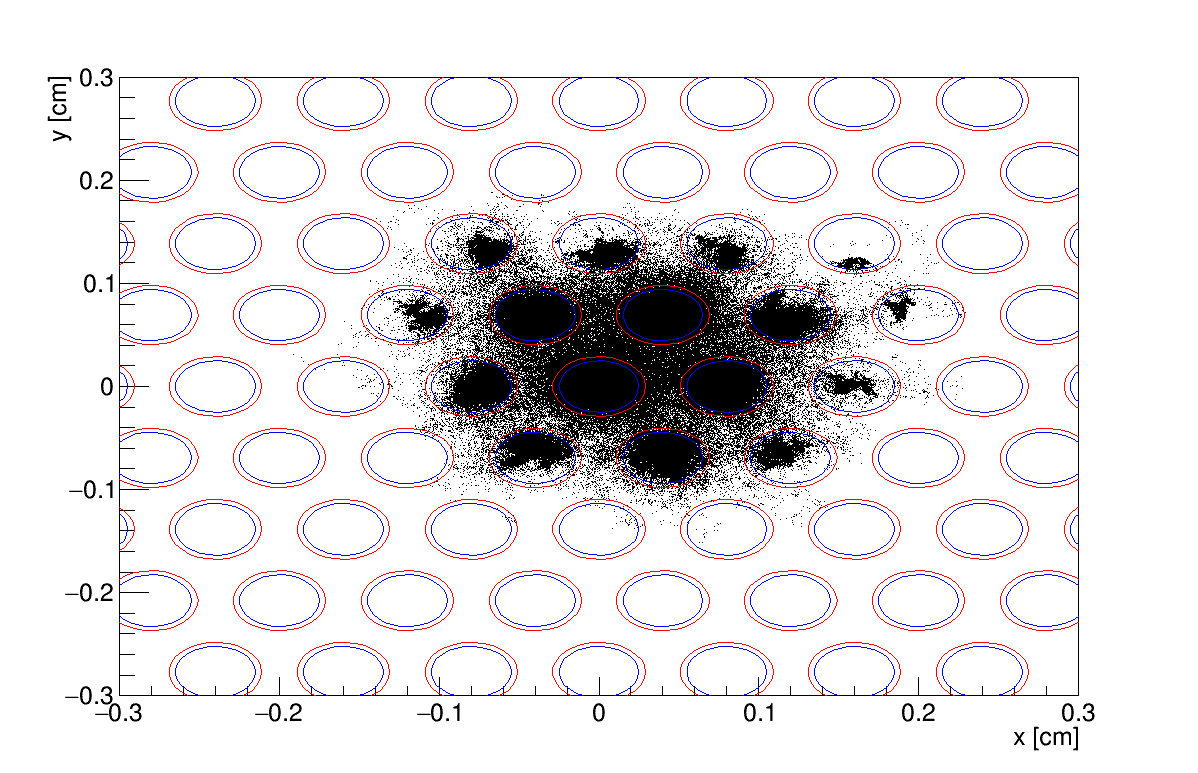}
\caption{Photons}\label{fig:XYPhotons}
\end{subfigure}
\caption{The event distribution in the $xy$ plane, perpendicular to the liquid surface, for the ionizations (a) and the photons (b). Black points in (a) indicate single ionization events, while the red points indicate the centers of the avalanches. The plots show the accumulation of 500 events at 3300 V across the LEM. A correlation between the two processes is clearly visible although while the ionizations are limited to the holes, the photon production is much more spread out. To guide the eye also the holes are drawn in the relevant region: the hole in the dielectric in blue and the one in the copper with a larger radius in red.}\label{fig:XYDistri}
\end{figure}

This is also reflected in Fig.~\ref{fig:XYDistri} which shows the production points in the $xy$ plane for charge and light. The
charge plot also shows the centre-of-mass of every avalanche as red points. In this projection, it is clearly visible, too, that
the ionization is occurring only within the holes, while the photon production is much more spread out.  However, a correlation between both is obvious. \\
The results for charge, light and the correlations between them are presented in the next sections. 

\section{Charge Avalanche}
\begin{figure}[ht]
\begin{center}
\includegraphics[width=0.8\textwidth]{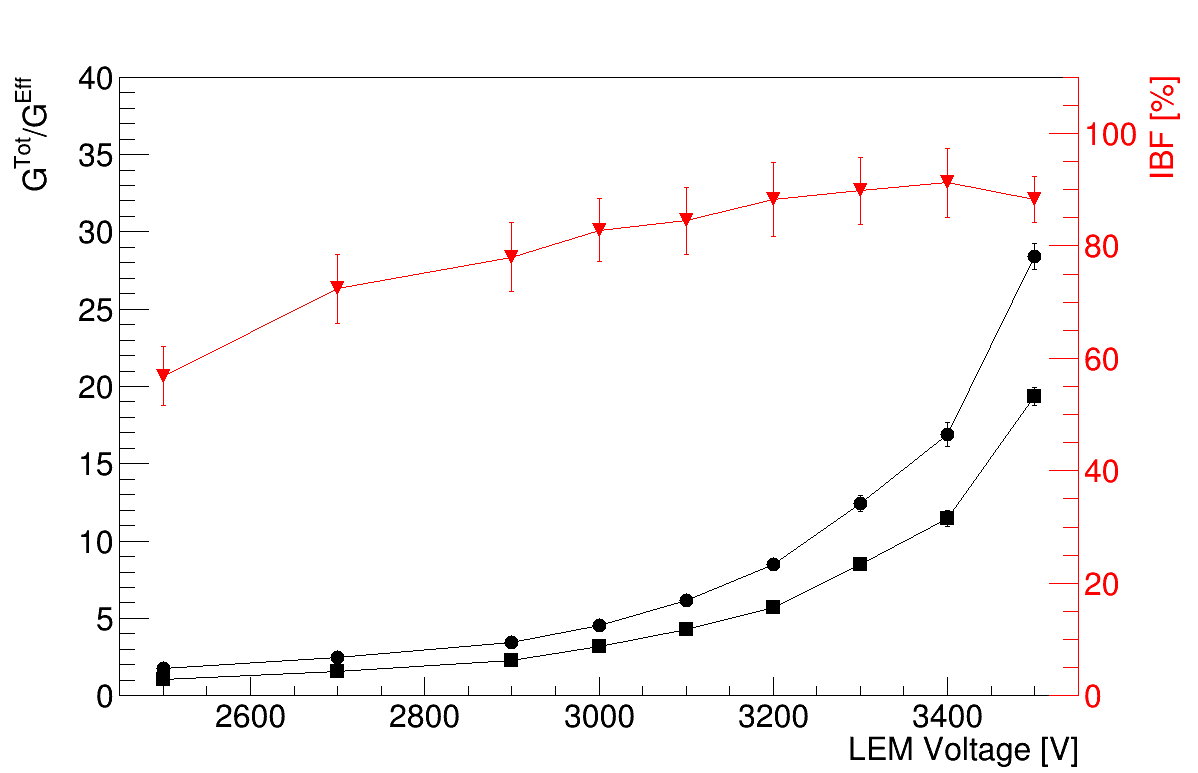}
\caption{Total (black dots) and effective gain (black squares) as a function of the applied LEM voltage. A high \ibf\ of about 60 \% for low voltages going up and saturating at about 90\% is observed for high LEM voltages. }
\label{fig:ChGainIBFVsLEM}
\end{center}
\end{figure}

The results for the \gtot, \geff\ and \ibf\ as a function of the LEM voltage are shown in Fig.~\ref{fig:ChGainIBFVsLEM}. For the total and the effective gain, an exponential rise with the LEM voltage can be observed. For a LEM voltage of 3500 V, an effective gain of 18 is obtained. The \ibf\, on the other hand, increases from 65\% at 2500 V to about 90\% at higher voltages where it saturates. The drop towards lower voltages comes from the fact that the ion produced together with the primary electron in the drift volume is not included. The result is not surprising comparing to experimental and simulation studies for GEMs with similar electric field configurations \cite{1221879, Bhattacharya:2017yaj}. Although the effective gain is low, the importance of the ion backflow should not be underestimated considering the high charge density along the track of a particle traversing the liquid argon. When operating on the surface, a perfectly vertical cosmic muon releases about 30.000.000 electrons 
over 6 m, which then drift towards the amplification region on the top. With a moderate effective gain of 11, around 300.000.000 ions will tend to drift back towards the liquid. While opposite opinions exist on whether if the ions enter the liquid \cite{Bueno:2007um, Romero:2016tla}, one has to consider that in the case that the ions enter the liquid, the field distortions in the extraction and drift regions and recombination effects due to these ions might be significant as shown in \cite{Romero:2016tla, 1748-0221-13-04-C04015}. As a consequence of the difference in drift velocity of electrons and ions of several magnitudes, in fact, the whole charge created by a 6 m long, vertical track, will be contracted to a sub-mm cloud of positive charges.  

\section{Secondary Scintillation}
\begin{figure}
\begin{subfigure}[b]{.5\linewidth}
\includegraphics[width=0.9\textwidth]{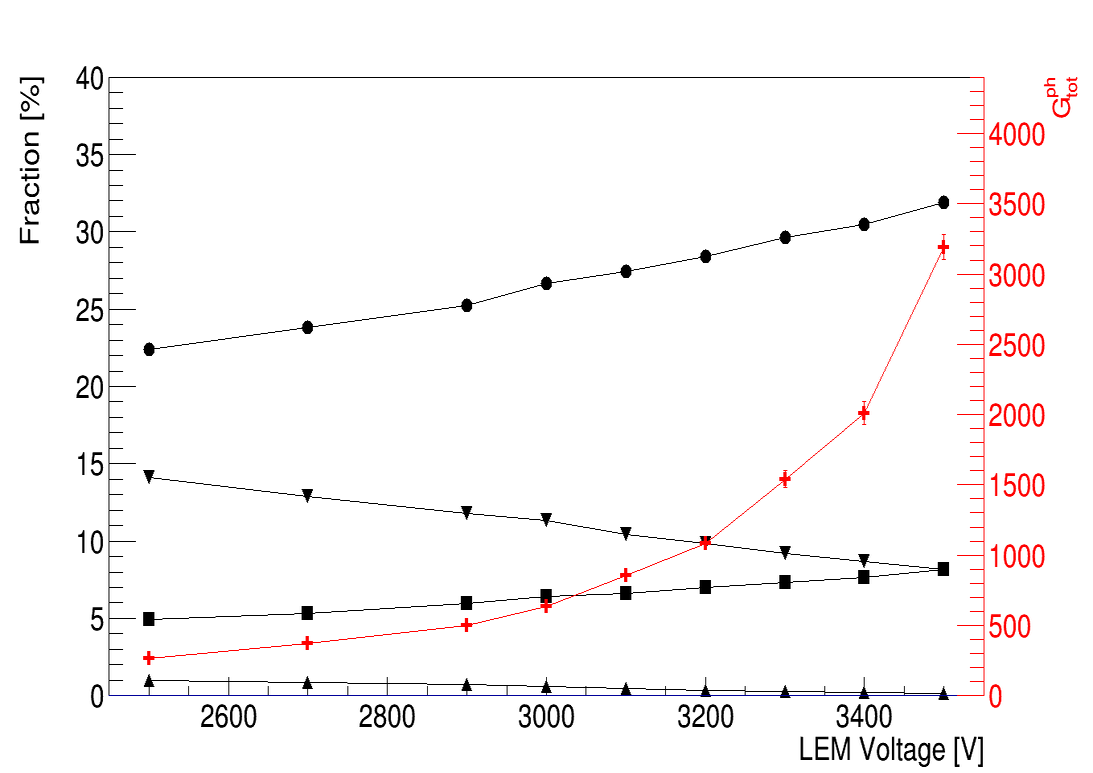}
\centering
\caption{\gphtot\ and photon end point distribution. }\label{fig:PhGainEnd}
\end{subfigure}%
\begin{subfigure}[b]{.5\linewidth}
\centering
\includegraphics[width=0.9\textwidth]{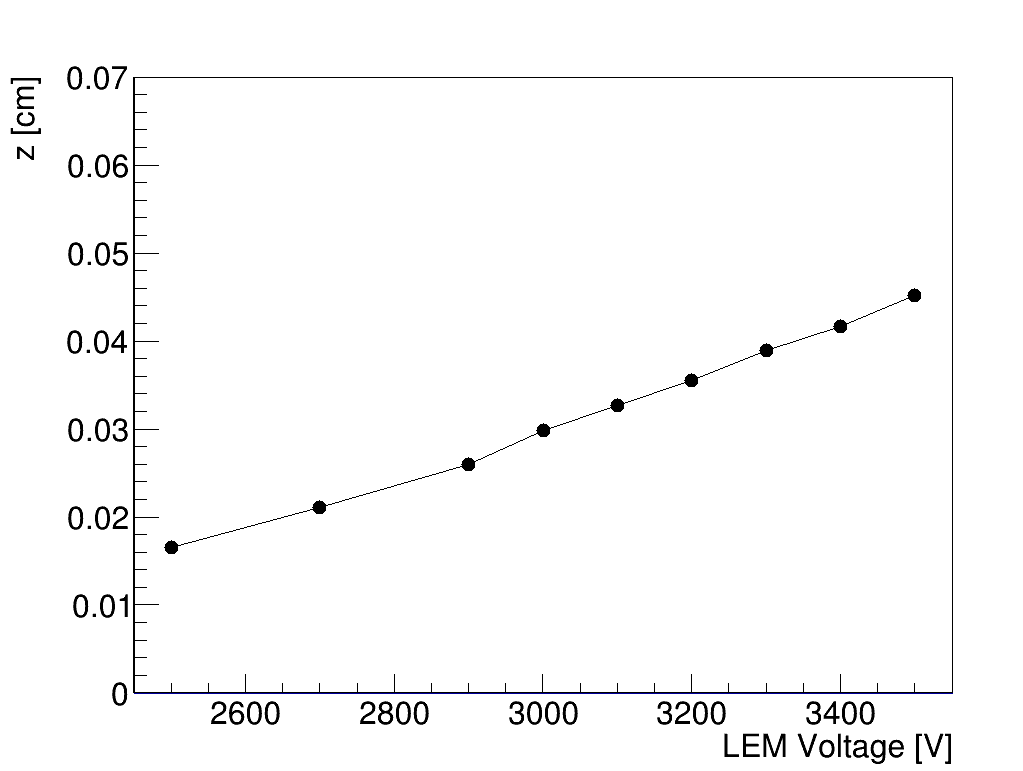}
\caption{$z$ mean production position versus LEM voltage}\label{fig:MeanZPhLEM}
\end{subfigure}
\caption{(a) The total photon gain as a function of the LEM voltage (red). Also the fractions of the total photons ending on one of the 3 electrodes, anode (dots) and top (squares) and bottom side (upward triangles) of the LEM or reaching the liquid argon surface (downward triangles) as a function of the LEM voltage. (b) Shifting of the mean $z$ position of the photon production in function of the applied LEM voltage.}\label{fig:PhEnd_ProdZ}
\end{figure}

The results for the secondary scintillation photons are shown in Fig.~\ref{fig:PhGainEnd}.  High light gains with LEMs were reported experimentally \cite{MONTEIRO201218} already for a different LEM geometry and field configuration and the results of this study with a total number of produced photons exceeding 3000 per primary electron at 3500 V agrees with that ones,
Each of the simulated photons was propagated through the detector and its end point was recorded. Using this information for each surface, the anode, LEM top side, LEM bottom side and liquid, the fraction of photons ending on each of these surfaces was determined. 
As one can see with increasing LEM voltage a higher fraction of photons end on the anode and the top side of the LEM although in absolute numbers too, for the bottom side and the liquid surface, more photons are measured with increased LEM voltage. This behaviour can be explained by the fact that the mean position of the photon production shifts towards the anode side with increasing LEM voltage. A similar result was reported in other MPGD studies \cite{1748-0221-7-09-P09006}

\begin{figure}
\begin{subfigure}[b]{.5\linewidth}
\includegraphics[width=0.9\textwidth]{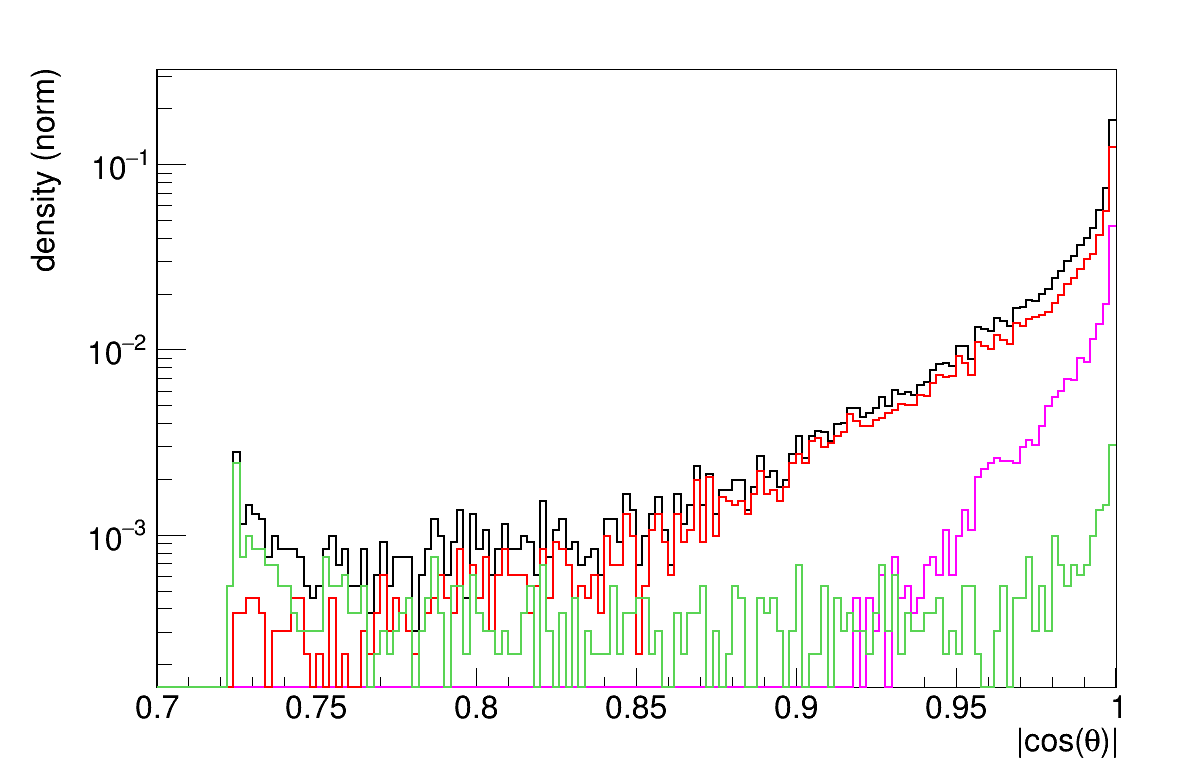}
\centering
\caption{Opening angle contributions}\label{fig:ThetaLContri}
\end{subfigure}%
\begin{subfigure}[b]{.5\linewidth}
\centering
\includegraphics[width=0.9\textwidth]{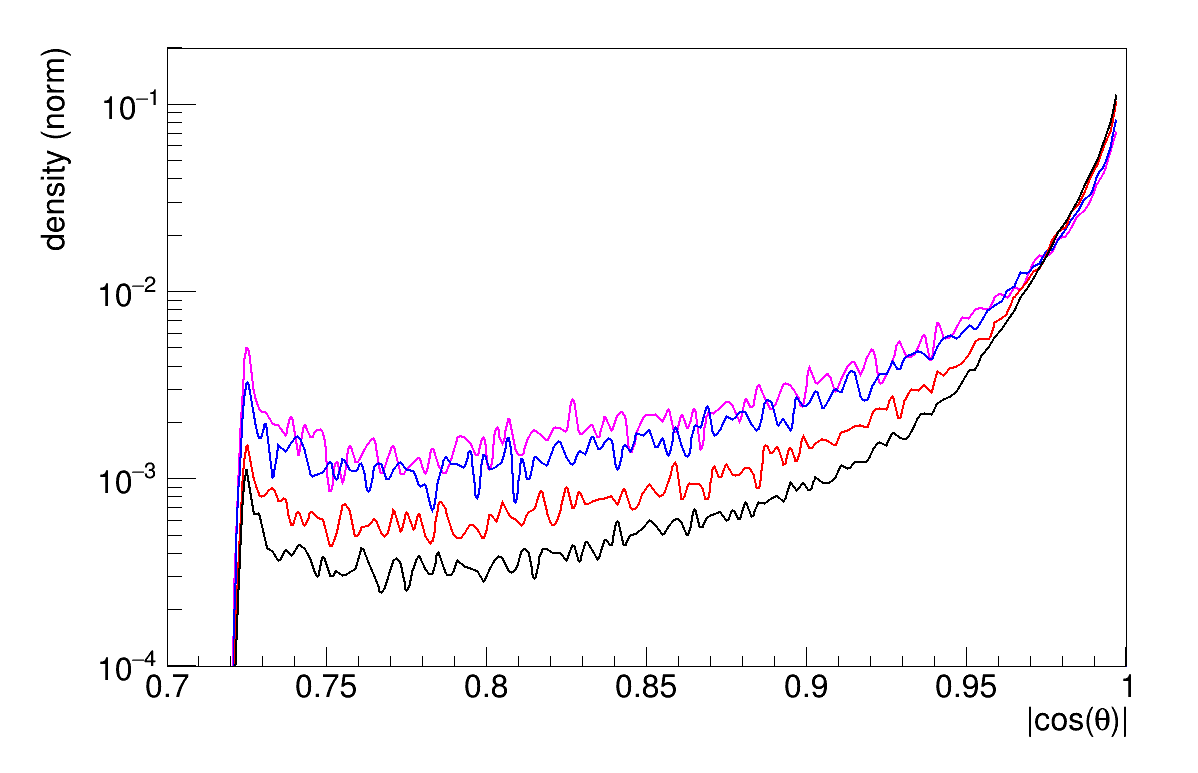}
\caption{Opening angle relation to LEM voltage}\label{fig:ThetaLLEM}
\end{subfigure}
\caption{(a) Absolute value of the cosine of the opening angle of the photon cone inside the liquid argon assuming a wavelength of 128 nm, \ngas= 1.0005 and \nliq= 1.45. Also the distributions of the different photon productions regions is shown: photons from the extraction region (blue), from inside the holes (red) and from the induction region (magenta). (b) The same distribution for different LEM voltages: 2500 V (magenta), 2900 V (blue), 3300 V (red) and 3500 V (black). }\label{fig:CosTheta}
\end{figure}

For the performance studies of a DUNE far detector based on the DP technology and possible future and optimization of the readout region, not only the amount of photons produced is relevant but the spatial distribution of the photons entering the liquid is especially interesting. Fig.~\ref{fig:CosTheta} shows the opening angle, the angle between the photon direction and the vector perpendicular to the liquid surface, in the liquid under the assumption of a refractive index of 1.0005 in gaseous and 1.45 \cite{LArProp} in liquid argon. Fig.~\ref{fig:ThetaLContri} shows the absolute value of $\cos \theta$ and the contribution to the distribution coming the from the different regions. The photons produced in the holes are the dominant contribution, while the photons produced in the extraction region add a long tail at large angles and the photons from the induction gap enhance the peak around 1. The dependence on the LEM voltage is shown in Fig.~\ref{fig:ThetaLLEM}. As expected the relevance of the photons from the extraction regions decreases with increasing LEM voltage, while the central region is increasing.  

\section{Correlations between Charge and Light}
\begin{figure}
\begin{subfigure}[b]{.5\linewidth}
\includegraphics[width=0.9\textwidth]{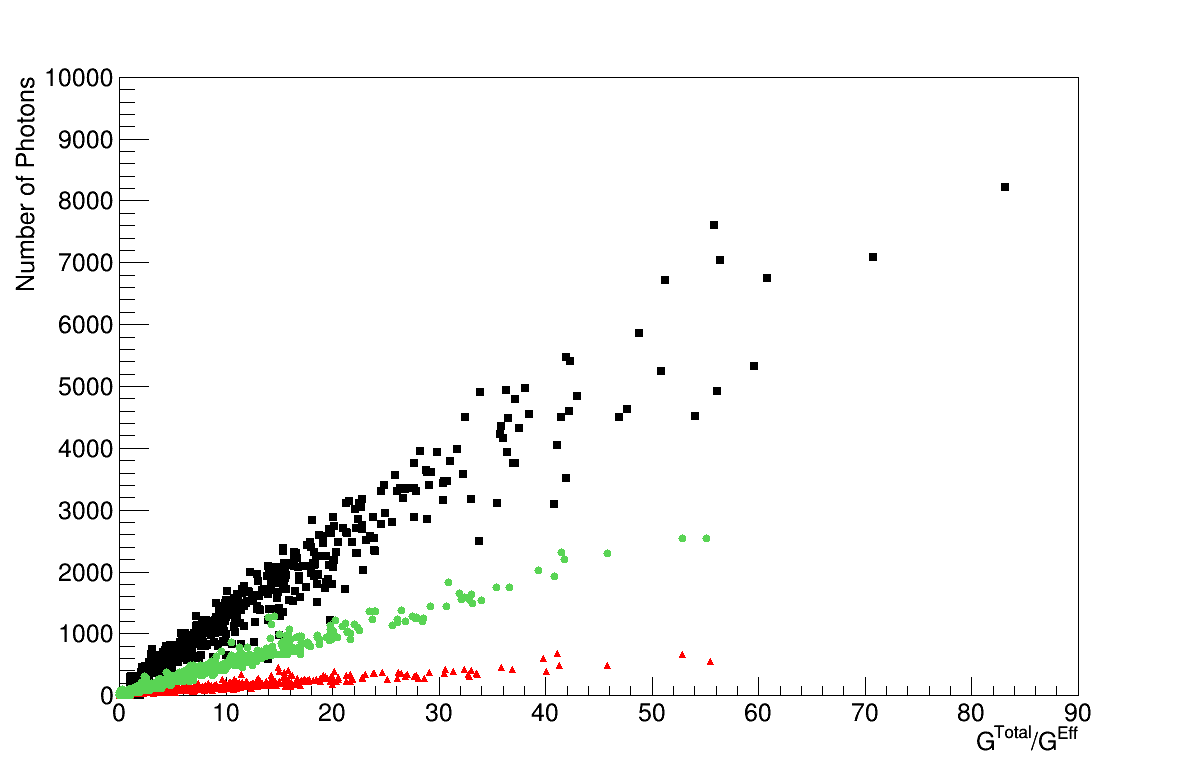}
\centering
\caption{Photon versus charge gain}\label{fig:ChVsPhG}
\end{subfigure}%
\begin{subfigure}[b]{.5\linewidth}
\centering
\includegraphics[width=0.9\textwidth]{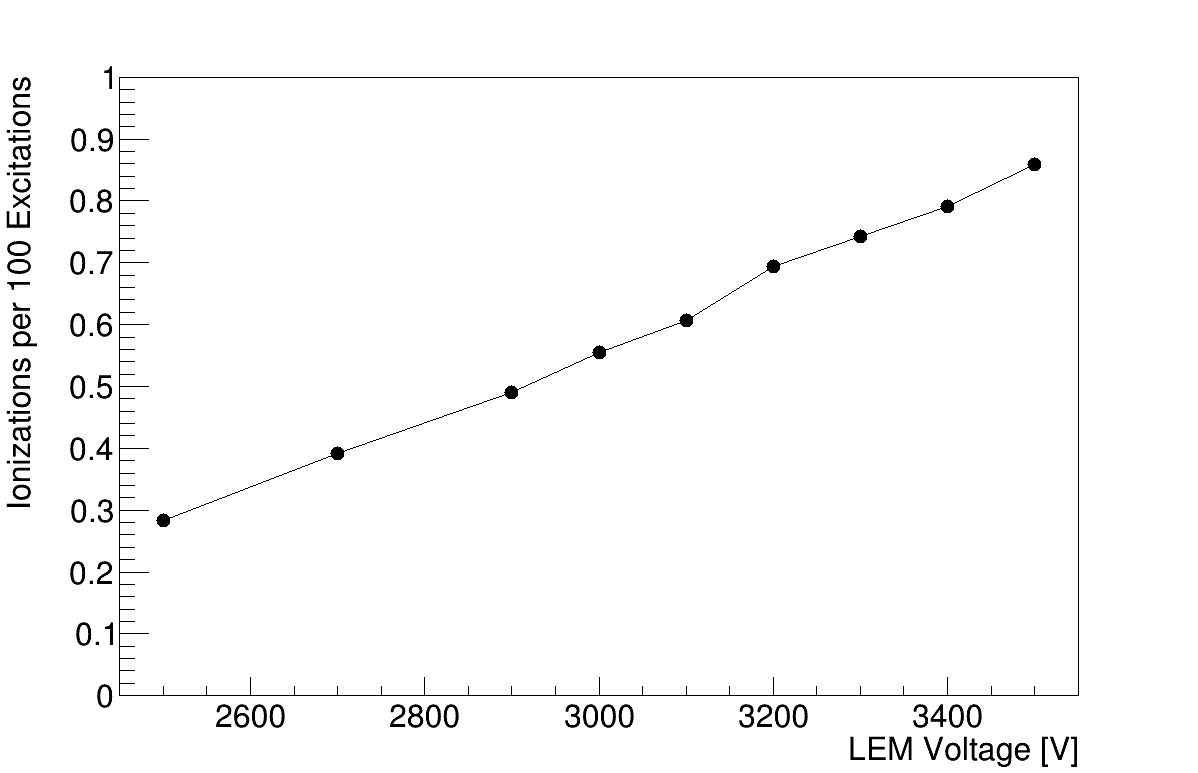}
\caption{Number of ionizations per 100 excitations.}\label{fig:RatIonExc}
\end{subfigure}
\caption{(a) Photon versus charge gain: \gphtot\ versus \gtot\ (black) and number of photons reaching the anode (green) and number of photons reaching the liquid (red) in function of \geff. 
(b) The number of ionizations per 100 excitations/photons. Less than 1 \% of the interactions
lead to ionization. }\label{fig:PhChCorr}
\end{figure}

The charge and the light production are strongly correlated due to the underlying physics process. Fig.~\ref{fig:ChVsPhG} shows the correlation for this study. The $x$ axis represents the charge gain, either the total or the effective gain. In black is shown the total number of photons in function of the total number of electrons produced in the same event. The figure also shows the number of photons reaching the anode (green) and the number of photons reaching the liquid (red) in function of the effective gain, the number of electrons reaching the anode. A more general study of the correlation can be found in \cite{Monteiro_2014}. Excitation and ionization are competing processes, with a clear dominance of excitation as shown in Fig.~\ref{fig:RatIonExc}. This is also reflected in the distributions shown in Fig.~\ref{fig:ZDistri}. As a function of the applied LEM voltage the fraction of ionization increases towards higher LEM voltages as expected.

\section{Photon Feedback as possible additional Contribution to Charge Gain}
\begin{figure}[ht]
\begin{center}
\includegraphics[width=0.6\textwidth]{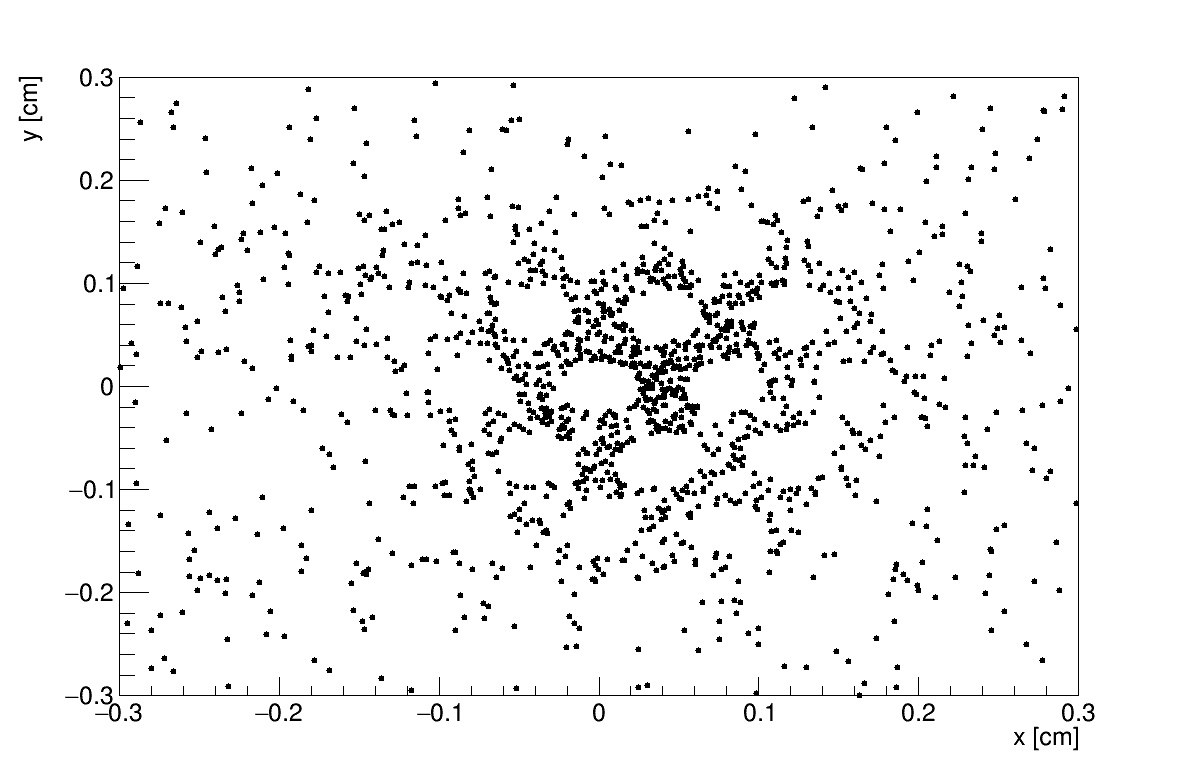}
\caption{Distribution of the photons ending on the bottom of the LEM for 3300 V for 500 events. }
\label{fig:S2ExtrBotLEM}
\end{center}
\end{figure}

While the results presented here on the dependencies of the charge and light production on the LEM voltages and the correlations between them seem to be reasonable and follow the tendencies reported by other studies e.g. \cite{1748-0221-7-09-P09006}, it has to be pointed out that the effective gain obtained in this study is significantly lower than the experimental results reported in \cite{Cantini:2014xza} for the same geometry and operation parameters. Considering that charging up effects were not simulated, the result of this study, 8.5 at 3300 V across the LEM, has to be compared to the reported gain of about 65 for the same voltage setting before the charging up, which reduces the effective gain for the same LEM voltage by a factor 1/3.3.\\
One difference between simulation and experiment is that the simulation was performed for pure argon, while small amounts of impurities are unavoidable for real argon gas. However, the usual impurities in laboratory environments like nitrogen, oxygen and water vapour, tend to reduce the gain for a given voltage but allow to achieve higher maximal gains by increasing the voltages. The gain limitations for THGEMs in pure noble gases were studied in detail for various pure noble gases in \cite{Miyamoto_2010}.
Temperature and pressure also affect the gain due to changes in the gas density which temperature and pressure variations imply but under normal ambient conditions this can only explain the gain difference at the percent level.\\
In fact, comparison of the results from Fig.~5 of \cite{Cantini:2014xza} with the results of this study indicates that the experimental data differs significantly from the expected exponential shape. This "over-exponential" behaviour can be also found in \cite{MONTEIRO201218} for argon as also to smaller extent for xenon. 
This suggests that in the experimental data additional gain production processes are involved, which are not included in the simulation. Photon feedback is a natural candidate for this additional charge gain.
Photon feedback is a well known process reported already in studies in the 80s and 90s \cite{FONTE199191, VAVRA1997137,doi:10.1063/1.327395}, including in pure argon. This is not surprising considering that the energy of the S2 photons in pure argon is between 8 and 12 eV ($\pm 3 \sigma$)  corresponding to 128 $\pm$ 10 nm, far above the work function for copper which for vacuum was measured to be about 4.5 eV \cite{PhysRev.76.388}. In this process, an electron is accelerated in a strong electric field to such high energies that it excites gas atoms by energy transfer in collisions. In the following de-excitation process a photon is produced which impinges on a photo-sensitive surface ejecting another electron which under the suitable circumstances undergoes the same process as the primary electron. As long as the condition for a positive feedback:
\begin{equation}
    1 > pG
\end{equation}
is fulfilled, with $p$ the probability that a secondary particle is produced and $G$ the gain, the process will not be completely unstable. However, a significant sensitivity to external parameters, as for example gas quality and field settings, can be expected. With this assumption and following the argument for the feedback loop gain from \cite{FONTE199191}, the overall charge gain can be written as:
\begin{equation}
    G_{Ch}^{Tot} = G_{Ch}^{Av} + \epsilon_{ph2ch} \cdot \frac{G_{ph}}{1-pG_{ph}}
\end{equation}
Here, $\epsilon_{ph2ch}$ is the efficiency with which photons are converted to electrons and $G_{Ch}^{Av}$ the 
charge gain due the avalanche process. As shown in Fig.~\ref{fig:ChGainIBFVsLEM} and Fig.~\ref{fig:PhGainEnd} both charge and light gain follow an exponential shape. It is worth studying the 2 extreme cases: a LEM voltage too low to produce a charge avalanche but high enough to excite the gas atoms and the case where the voltage is very high. In the first case, the observed gain would be above 1 due to the contribution of the photon feedback coming from the primary electron. In the latter case, the measured gain would rise faster than exponential due to the gain loop from the positive feedback assuming that the change in the field configuration does not make the efficiency to convert photons in electrons vanish. \\
Under the assumption that photon feedback is occurring in the experimental setup, it is interesting to analyze where it could appear, which photon production regions and surfaces could be involved and which are the implications on the detector performance for the different cases:\\
The photons produced in the holes can only impinge on the anode or enter the liquid. In the first case the electrons will be directly pulled back and in the latter case, they will leave the interesting region by entering the liquid argon to be detected partly by the PMTs installed behind the highly transparent cathode at the bottom of the detector. \\
In the extraction region, only a few photons are produced, especially in the vicinity of the hole opening (Fig.~\ref{fig:S2ExtrBotLEM}), but due to the high electric fields created by the high voltage across the LEM on the bottom LEM surface the probability to extract an electron and guide it to the hole can be highly enhanced in this region.
The electrons released there will undergo the same avalanche process as the primary electron, doubling not only the charge gain but also the \ibf\ and the photon gain. 
The same process is exploited in gaseous photo detectors based on LEMs \cite{BRESKIN2011117} for RICH detector applications.\\

\begin{figure}
\begin{subfigure}[b]{.5\linewidth}
\includegraphics[width=0.9\textwidth]{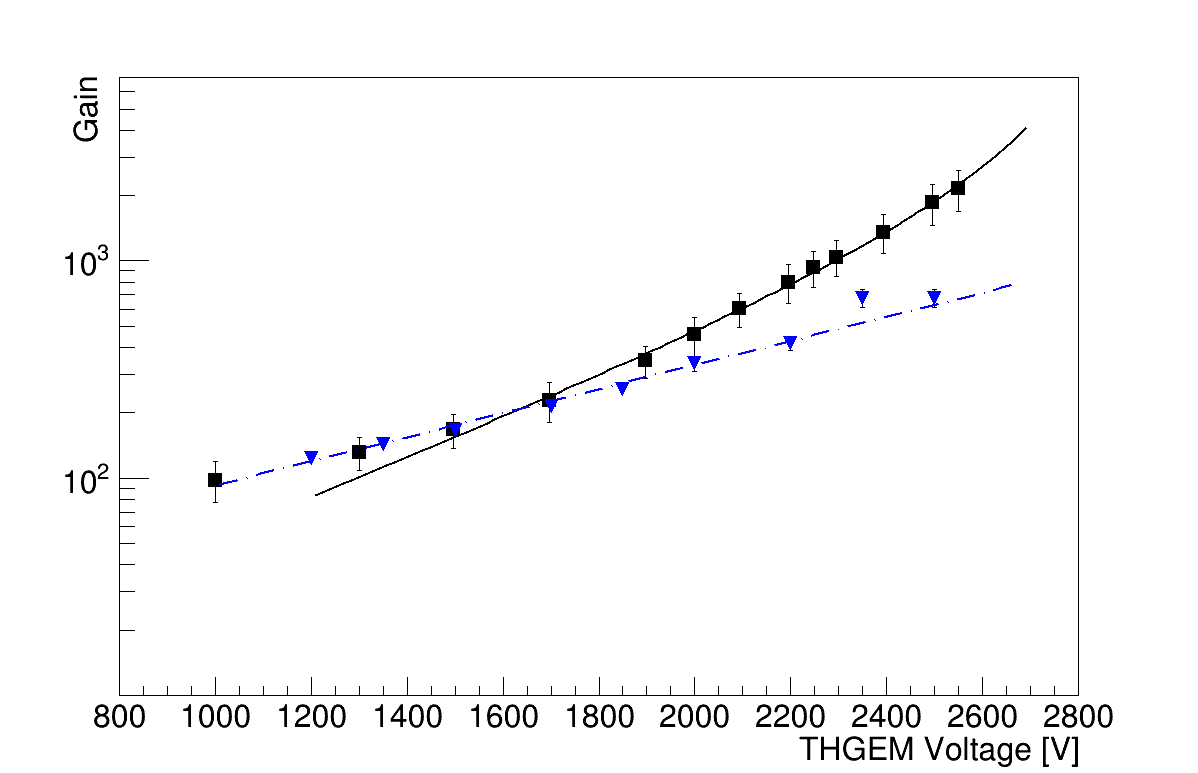}
\centering
\caption{Xenon data at 2.5 bar}\label{fig:XeCoim}
\end{subfigure}%
\begin{subfigure}[b]{.5\linewidth}
\centering
\includegraphics[width=0.9\textwidth]{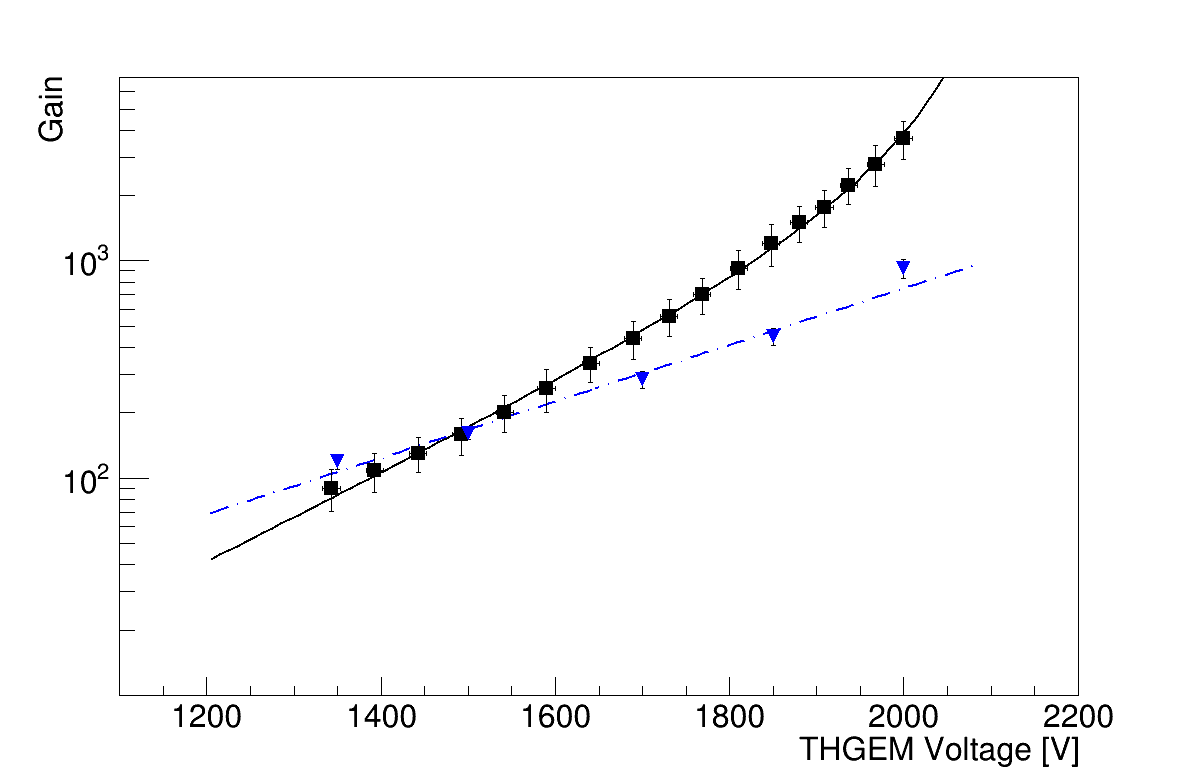}
\caption{Argon data at 2.5 bar}\label{fig:ArCoim}
\end{subfigure}
\caption{Direct comparison of gain versus voltage across THGEM of data (black squares), extracted from \cite{MONTEIRO201218}, with simulation results (blue triangles): (a) for xenon at 2.5 bar and (b) for argon at 2.5 bar. The experimental data was fitted with a positive feedback gain (solid line) and the simulated data with an exponential (dashed line). }\label{fig:DataCoimbrar}
\end{figure}

In the induction region, significantly more photons are produced, partly due to the higher electric field strength compared to the extraction region, partly due to the larger number of electrons available after the charge avalanche within the holes. This contribution will depend strongly on the field strength in this region since only photons absorbed in regions of the top electrode of the LEM where field lines towards the anode start can contribute. As soon as $E_{ind}\leq E_{ex}$ the photon feedback gain should vanish. Photon feedback in this region would only increase the charge gain but no more ions would be produced. As shown above, also most of the light produced by the released photons would not reach the liquid and therefore not spoil any S1 measurements. \\
In the following these considerations are applied to the experimental results of 3 different studies: the previously mentioned electroluminescence study with pure argon and xenon using a LEM/THGEM \cite{MONTEIRO201218}, a study about the effect of the rim size presented also in \cite{Cantini:2014xza} and a study of the performance of a single GEM in pure argon \cite{BRESSAN1999119}. \\
In \cite{MONTEIRO201218} one can see an over-exponential shape in the light gain curve. The drift field is quoted as 0.5 kV/cm and the induction field was set in the range of 2 to 4 kV/cm without specifying which field was used for which data set. If one assumes the lower limit, the clear difference from the exponential behaviour is surprising under the assumption that photon feedback is responsible for the over-exponential shape of the gain curve. For this reason, the geometry and field configuration used in that study was simulated at 2.5 bar using the same code developed for the results presented previously. The results are shown in Fig.~\ref{fig:XeCoim} for xenon. For low voltages, the simulated and experimental results agree well for voltages below 1600 V but both curves diverge for higher voltages. An exponential gain with positive feedback was fitted to the experimental data in the range of 1400 to 2550 V:
\begin{equation}
    G^{meas}_{ph} = a\frac{e^{bV}}{1-p\cdot e^{bV}}
\end{equation}
The curve describes well the data, except for low voltages. Considering that the probability, $p$, of the gain feedback depends on various factors as for example the probability to eject an electron from the copper and the efficiency 
to guide the electron afterwards into the LEM hole which are both functions of the LEM voltage, it is obvious that a fit assuming 
a constant $p$ cannot describe the full data range because for low voltages the photon feedback should vanish and the curve should follow
an exponential. The simulated data were fitted with an exponential. Fig.~\ref{fig:ArCoim} 
shows the results for argon. The fit for the experimental data, which was extracted from the publication using suitable tools\footnote{www.DataThief.org}, was performed in the range of 1300 to 2200 V. Also, in this case, a significant difference
between experimental and simulated data is found, both in the size of the gain as in the shape of the gain curve. On the other hand the 
hypothesis of photon feedback can not explain that the appearance that the simulated results lay slightly over the experimental data for low voltages. 
However, experimental explanations might be found for this because e.g. impurities tend to quench the S2 production and to absorb the photons of 128 nm, also an overestimation of the quantum efficiency to 128 nm of the used light sensor could contribute to this observation. Analyzing the simulated argon data further it was found, that the electric field at 1700 V is leaking about 300 $\mu$m into the drift field and about 400 $\mu$m into the induction gap yielding about 15 and about 50 photons respectively in these regions. The photons from these regions might be responsible for photon feedback yielding the observed data curve. \\
The fact that the same effect is observed not only in argon but also in xenon, and there with a smaller divergence, can be attributed to the 
fact that also in xenon scintillation light is produced but with a lower photon energy of 7.2 eV (peak value). 

\begin{figure}[ht]
\begin{center}
\includegraphics[width=0.7\textwidth]{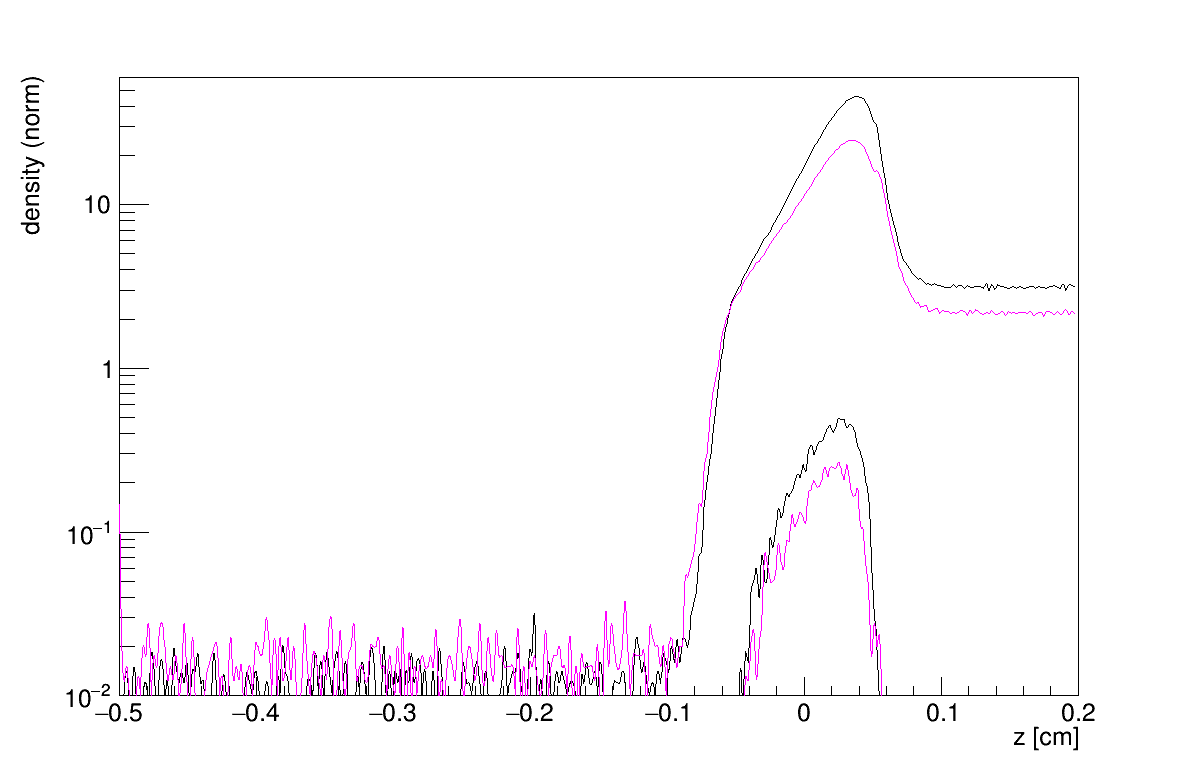}
\caption{Comparison of the production of photons and ionization along the $z$ axis for two different rim sizes: 40 $\mu$ 
 (black) and 80 $\mu$m (magenta). $z=0$\ cm corresponds to the center of the LEM.}
\label{fig:RimZ}
\end{center}
\end{figure}

Another case is the rim size, the additional radius in the copper hole, study presented in \cite{Cantini:2014xza}. For a rim of 40 $\mu$m a 4 to 5 times higher gain, depending on the applied LEM voltage, than for a rim of 80 $\mu$m was measured. 
To compare these results with the simulation, the same configuration as before was simulated except for an increase of the rim size from 40 to 80 $\mu$m. It was found that both charge and photon gain drop by about 40\%. The effect on the production along $z$ is shown for charge and light in Fig.~\ref{fig:RimZ}. Considering that the larger rim reduces also the probability of photon feedback, simply due to solid angle effects, a combination of charge and photon gain reduction might
be sufficient to explain the observed large drop in gain as well as the change in the shape of the gain curve towards an exponential.\\

\begin{figure}
\begin{subfigure}[b]{.5\linewidth}
\includegraphics[width=0.96\textwidth]{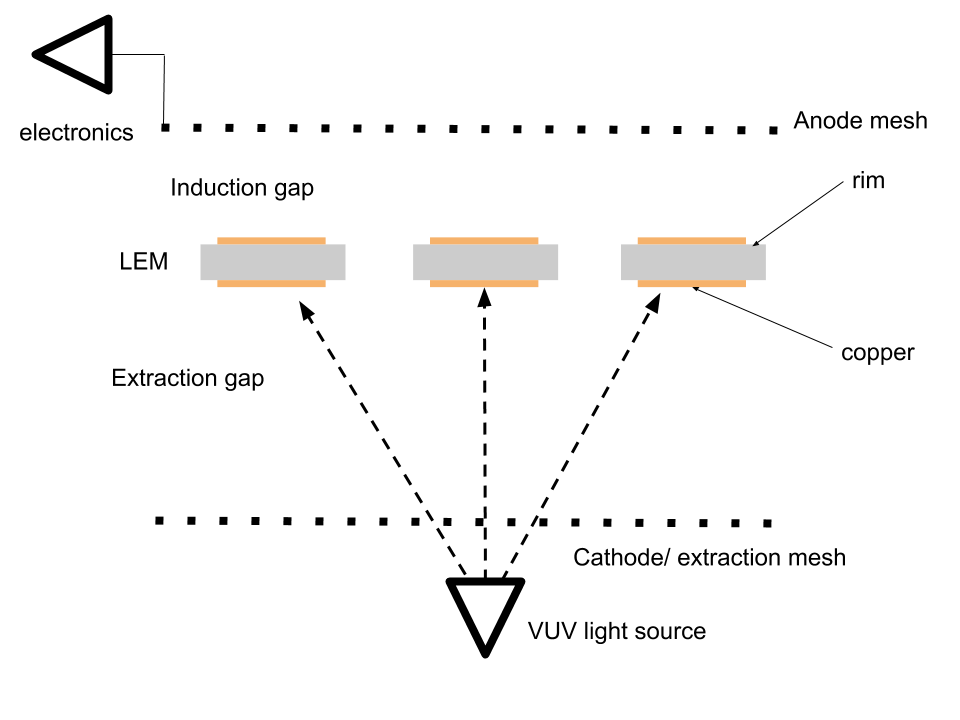}
\centering
\caption{Photon effect on the bottom side of the LEM}\label{fig:ExpSketchExt}
\end{subfigure}%
\begin{subfigure}[b]{.5\linewidth}
\centering
\includegraphics[width=0.96\textwidth]{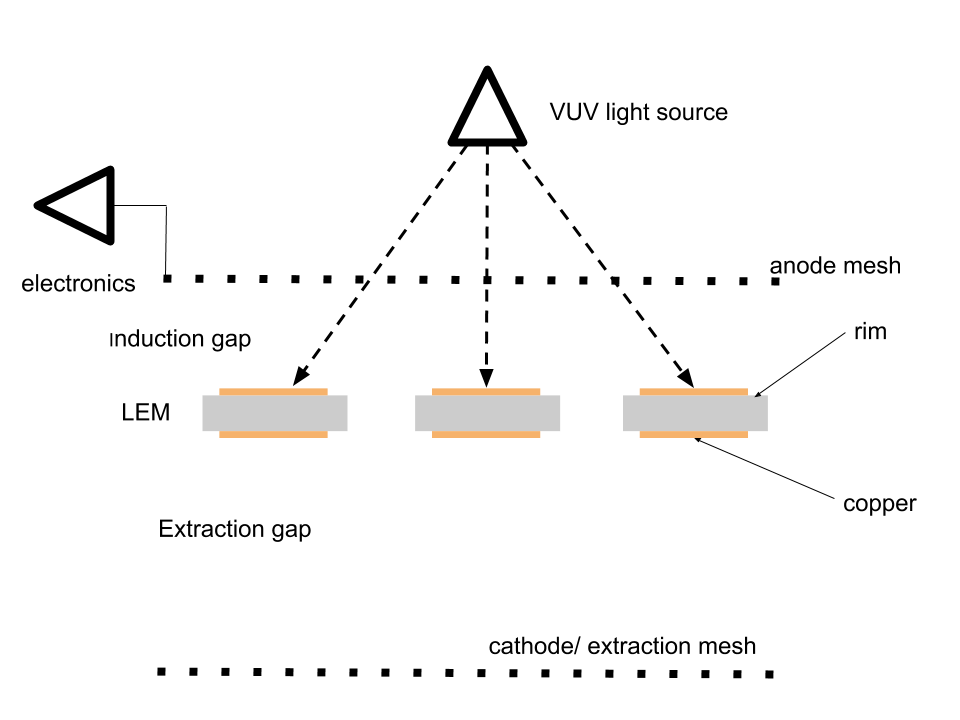}
\caption{Photon effect on the top side of the LEM}\label{fig:ExpSketchInd}
\end{subfigure}
\caption{Sketch of a possible setup to study separately the photon feedback effect: (a) for the extraction
gap and (b) for the induction gap.}\label{fig:setupSketch}
\end{figure}

High gains in pure argon were not only reported for LEMs but also for GEMs \cite{BRESSAN1999119}. This study is interesting for two reasons: They report a strong dependence of the effective gain on the electric fields above and below the GEM as one would expect for feedback effects due to photons produced in these electric fields. Furthermore, they compared the gain measured in pure argon with the one measured in the same setup with an Ar:CO$_2$ mixture. While the gain curve for pure argon diverges from the exponential curve, as discussed above, the curves for
the gas mixture show the expected exponential behaviour. Considering that CO$_2$ efficiently absorbs VUV
photons, these two measurements can be seen as one measurement with possibly includes photon feedback effects and avalanche charge gain and one measurement exploiting only the charge avalanche process.\\
While one might interpret these considerations as a hint that photon feedback effects contribute to the reported gains in pure argon detectors, to prove or disprove this hypothesis additional experimental data with dedicated setups will be necessary. There, it would be important to measure not only the overall relevance of this effect but to distinguish between the contributions from the extraction and the induction gap due to the different impact on the overall detector performance. A possible experimental approach is sketched in Fig.~\ref{fig:setupSketch}. Important would be
to use the same geometry and electrical settings as in the LAr TPC. Instead of cryogenic temperatures a pressure of about 3.2 
bar could be used to have the same gas density. The cathode and anode should be made of highly transparent meshes behind
which a strong VUV light source, a xenon lamp or an electrically decoupled electroluminescence detector, would be installed depending
on whether the photon feedback in the extraction or the induction gap is studied. Measuring the charge signal on the anode mesh
would provide information about the size of the photon feedback effect. 

\section{Conclusions}
The paper reports the results of a simulation study about the charge and photon production in the charge readout region as planned for the future DUNE far detector. 
The results agree qualitatively with the expectations based on similar studies
for other configurations. The ion feedback is predicted to be very high, in agreement with studies carried out
for GEMs for similar electric field configurations, and might lead to severe field distortions, especially for the case of
surface operation of such detector. The charge and photon gain was found to be highly correlated showing an exponential behaviour.
It was found that a significant amount of S2 photons will enter the liquid and might affect the trigger capability of the detector
on no-beam physics. As input for the performance studies of the future DUNE far detector the spatial distributions of the photons entering the liquid are provided, important together with the number of photons for evaluating the trigger capabilities of a DUNE far detector based on the DP LAr technology.\\
Significant differences were found between simulated gain and experimental data from various studies, not only in the magnitude
of the gain but also in the shape of the gain curve. The over-exponential 
shape of the gain curves of the experimental data could be interpreted as the presence of additional charge gain processes,
beyond the common avalanche process. It is suggested that photon feedback effects could be responsible for this considering the
high energy of the S2 photons produced in argon and in nobles gases in general. Based on experimental
studies carried out over the last decades and considering the high energy of argon scintillation light it can be assumed that the effect occurs in the conditions considered here. \\
However, whether the effect is large enough to explain the difference between the simulated and experimental data, can only be proved 
with additional measurements aiming at explicitly studying this effect. If the hypothesis is confirmed that positive
feedback due to S2 photons contributes significantly to the observed charge gain, it might open the opportunity to develop optimized readout schemes for the DUNE DP far detector with reduced ion backflow and backwards going S2 light. Promising alternatives to LEMs for this optimization are already developed and tested with small size detectors \cite{VELOSO2011134}.
 
\acknowledgments
The author would like to thank Filippo Resnati, Federico Sanchez, Matteo Cavelli-Sforza and the members of his research group for the 
discussions related to this publication as well as the proof reading. He also would thank the Garfield/Magboltz team around Rob Veenhof, Heinrich Schindler, Steve Biagi and especially Josh Renner for providing the framework which was used in this study. \\
This project has received funding from the Spanish Ministerio de Econom$´{i}$a y Competitividad (SEIDI-MINECO) under Grants no.~FPA2016-77347-C2-2-P and SEV-2016-0588.

\bibliographystyle{JHEP}
\bibliography{biblio}

\end{document}